\documentclass{article}
\usepackage{cite}
\usepackage{graphicx}
\usepackage{arxiv}
\usepackage{svg}
\usepackage[space]{grffile}
\usepackage{hyperref}
\usepackage{todonotes}

\title{Hamiltonicity in Semi-Regular Tessellation Dual Graphs}

\author{
  Divya Gopinath \\ 
  \texttt{MIT} \\
  \texttt{divyagop@mit.edu} \\
  \And 
     Rohan Kodialam \\
     \texttt{MIT} \\
   \texttt{kodialam@mit.edu} \\
   \And
  Kevin Lu \\
  \texttt{MIT} \\
  \texttt{kezilu@mit.edu} \\
   \And 
  Jayson Lynch \\
  \texttt{MIT} \\
  \texttt{jaysonl@mit.edu}
   \And
   Santiago Ospina \\ 
   \texttt{MIT} \\
   \texttt{sospina@mit.edu} \\
}

\begin{document}
    \maketitle 
    
    \begin{abstract}

This paper shows NP-completeness for finding Hamiltonian cycles in induced subgraphs of the dual graphs of semi-regular tessilations. It also shows NP-hardness for a new, wide class of graphs called augmented square grids. This work follows up on prior studies of the complexity of finding Hamiltonian cycles in regular and semi-regular grid graphs \cite{itai1982hamilton,arkin2009not,hou_lynch}.

 \end{abstract}

    \section{Introduction}
    In this paper, we consider the problem of determining whether Hamiltonicity (finding a Hamiltonian cycle) is NP-complete in various grid graphs. The problem of finding a Hamiltonian cycle in grid graphs is a heavily-studied problem with a variety of applications in wireless network routing \cite{networks} and the lawnmower problem \cite{lawnmower}. We draw on work dating back to 1982, when Itai showed that the Hamiltonian path problem is NP-complete in square induced grid graphs \cite{itai1982hamilton}. More recent research has also shown that the Hamiltonian path problem is also NP-complete in other grids \cite{arkin2009not}, namely triangular and hexagonal grid graphs. These two results cover the three regular tilings of the plane by a single shape. \\ \\
    A natural extension to regular tilings are the family of semiregular tessellations, in which multiple polygons are are used to cover the plan. There are eight such tilings, and prior work has shown that Hamiltonicity is hard in these tessellations as well \cite{hou_lynch}. We continue this work by showing the Hamiltonian path problem is NP-complete in the \emph{duals} of all eight semi-regular tessellations, in which we create a vertex for each face of the tessellation and an edge between vertices corresponding to adjacent faces. We further consider the problem of Hamiltonicity in augmented square grids, in which each square cell can optionally contain one or two edges connecting the diagonals of the square. This class generalizes a number of prior results and includes important grids such as the king's move, the box-pleat, and the right triangle tessellation grids.

    \section{Problem Setup}
\begin{figure*}[ht!]
    \label{fig:semireg}
    \centering
    \includegraphics[width=130mm]{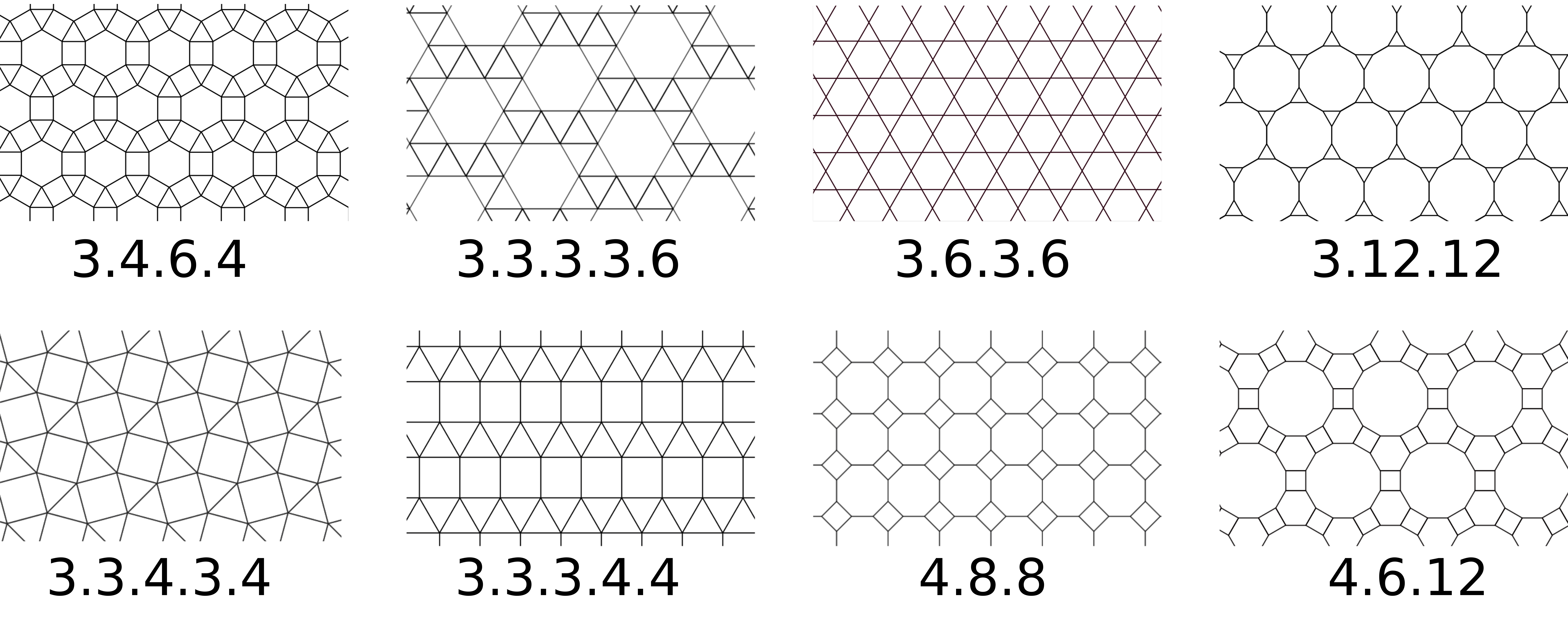}
    \caption{The eight semiregular tessellations}
    \label{fig:semi}
    \end{figure*}

  A \emph{tessellation} is a non-overlapping tiling of the plane with polygons. we consider two classes of tessellations: \emph{regular tessellations}, in which a single kind of regular polygon is used to tile the plane; and \emph{semiregular tessellations} formed by two or more regular polygons such that the same polygons in the same order surround each of the vertices formed by the tessellation. There are eight such tessellations, as shown in Figure~\ref{fig:semi} \cite{naturalstructure}. They are denoted by picking an arbitrary start vertex in the primal graph, and listing the size of the adjacent shapes while walking around a vertex in a counterclockwise fashion. For example, a vertex that is connected to a square and two octagons would be denoted as 4.8.8. \\ \\ 
    \begin{figure*}[ht!]
    \label{fig:dualsemireg}
    \centering
    \includegraphics[width=132mm]{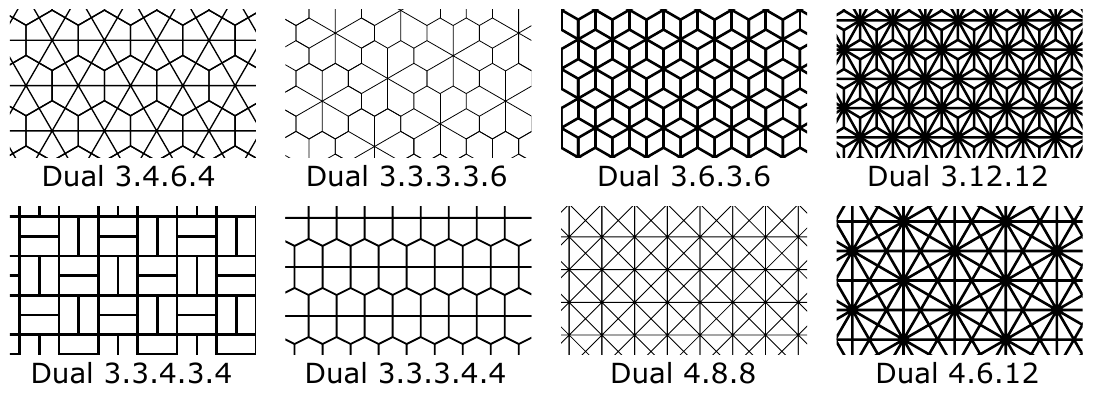}
    \caption{The duals of the eight semiregular tessellations.}
    \label{fig:dual_semi}
    \end{figure*}
    In this paper, we focus on the duals of these semiregular tessellations. Thus, for each tessellation $T$ we can define a graph $\mathcal{G}_T$ as follows: we let each polygon in $T$ correspond to a vertex in $\mathcal{G}_T$, and add an edge between two vertices in $\mathcal{G}_T$ if their corresponding polygons share an edge. \\ \\
    We define a \emph{Grid Graph} $G_T$ to be any induced sub-graph created by selecting an arbitrary subset of vertices of $\mathcal{G}_T$ and edges connecting vertices in this subset. We consider the problem of finding \emph{Hamiltonian Cycles} in these grid graphs, which are cycles that pass through each vertex of $G_T$ exactly once. Specifically, we consider the decision version of the Hamiltonicity problem which just determines whether or not the graph can admit such a cycle. \\
    A \emph{thin} grid graph is one in which every vertex of the sub-graph $G_T$ its corresponding vertex in the parent graph $\mathcal{G}_T$ has a neighbor which was not included in $G_T$. This definition differs from that of \cite{arkin2009not} which defines thin with respect to holes and pixels in the graph. However, this definition is functionally equivalent for the graphs under inspection, is easier to state, and avoids problems of trying to define pixels for augmented square grid graphs.
    \subsection{Approach}
    Generally, we use two main techniques in our reductions. First, we attempt to reduce Hamiltonicity in one grid graph to Hamiltonicity of another grid graph, usually a square or hexagonal one. In this paradigm, we seek to find either a subset of vertices whose induced subgraph is a simple grid, or to create vertex and edge gadgets that form an effective simple grid. These are in Sections~\ref{sec:simple} and \ref{sec:Hex}. \\ \\
    The second type of reduction we employ uses the Tree-Residue Vertex Breaking technique recently developed by Demaine and Rudoy \cite{demaine2017tree}. These can be seen in Section~\ref{sec:TRVB}. In this approach, we construct vertex gadgets that have open and closed configurations. These two techniques suffice for showing NP-hardness results for all $8$ of the duals of the semiregular tessellations.

    \section{Results}
    We show that Hamiltonicity in the grid graphs corresponding to the duals of all eight semiregular tessellations is NP-Hard. We divide our results into three sections. Section~\ref{sec:simple} examines simple reductions that follow directly from dual graphs. Section~\ref{sec:Hex} focuses on the creation of vertex and edge gadgets to prove NP-Hardness. Finally, Section~\ref{sec:TRVB} gives reductions from Tree-Residue Vertex Breaking.
    
    \subsection{Simple reductions}
    \label{sec:simple}
    In this section, we consider graphs where the induced grid graphs can emulate another grid graph by simply picking a subset of vertices whose induced subgraph is a regular tiling of the plane. 
    
    \subsubsection*{Dual of 3.6.3.6}
\begin{figure}[h]
    \centering
    \begin{minipage}[b]{0.45\linewidth}
        \centering
        \includegraphics[width=5cm]{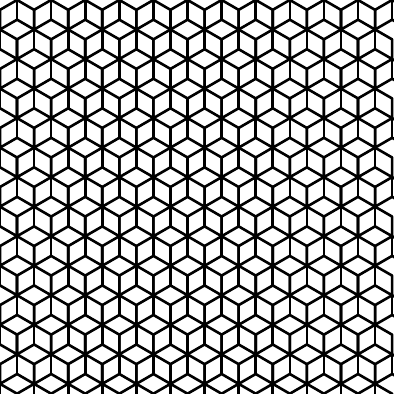} \\
        \caption{Dual of 3.6.3.6, also known as the rhombile tiling.}
        \label{fig:rhombile}
    \end{minipage}
    \hspace{0.5cm}
    \begin{minipage}[b]{0.45\linewidth}
        \centering
        \includegraphics[width=5cm]{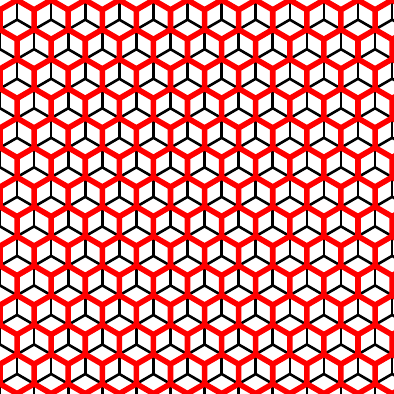} \\
        \caption{Red edges are an induced subgraph of the rhombile tiling and are a hex grid.}
        \label{fig:rhombile_hex}
    \end{minipage}
    \end{figure}

    The dual of 3.6.3.6 is known as the rhombile tiling, and contains the hexagonal grid as an induced subgraph. In Figure \ref{fig:rhombile}, the graph induced by the red vertices is a hexagonal grid graph. Since it is known that Hamiltonicity in hexagonal grid graphs is NP-hard, Hamiltonicity is also NP-hard in the rhombile tiling.

    \subsubsection*{Dual of 4.8.8}
\begin{figure}[h]
    \centering
    \begin{minipage}[b]{0.45\linewidth}
        \centering
        \includegraphics[width=5cm]{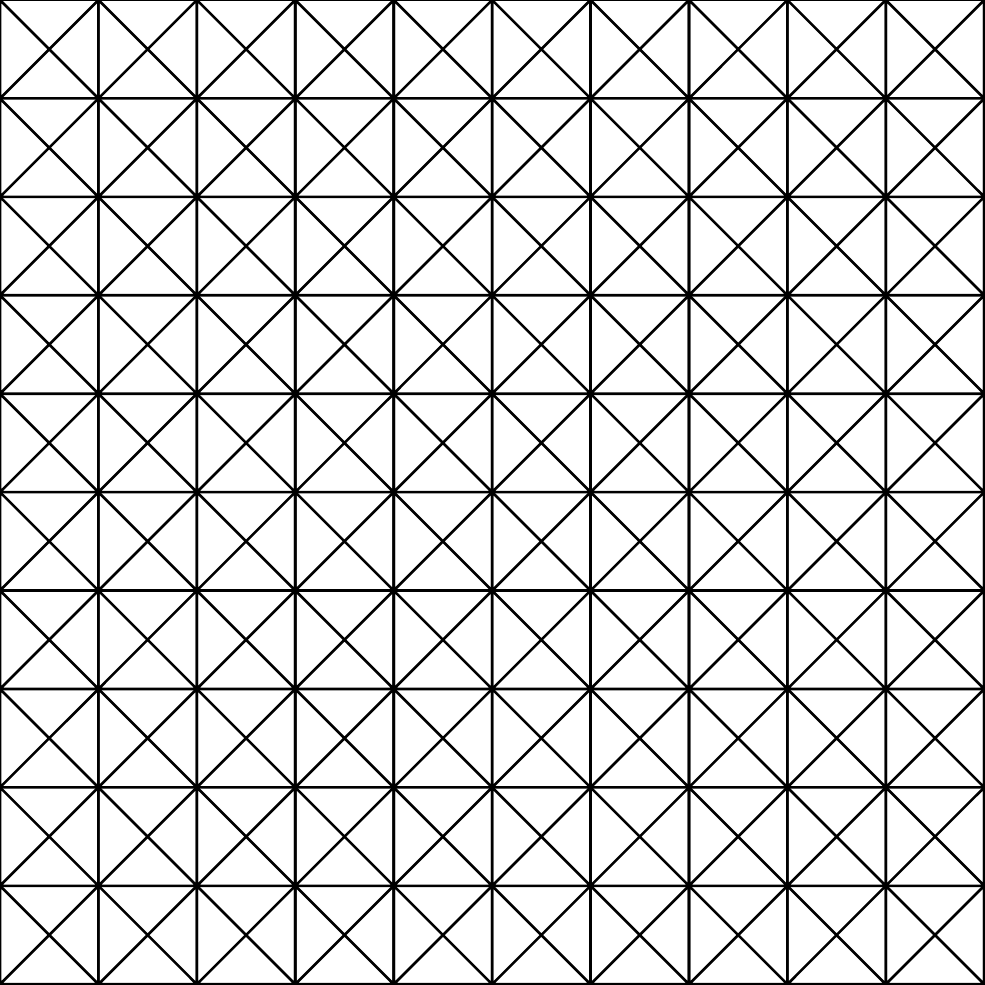}
        \caption{Dual of 4.8.8, also known as the Tetrakis grid.}
        \label{fig:tetrakis}
    \end{minipage}
    \hspace{0.5cm}
    \begin{minipage}[b]{0.45\linewidth}
        \centering
        \includegraphics[width=5cm]{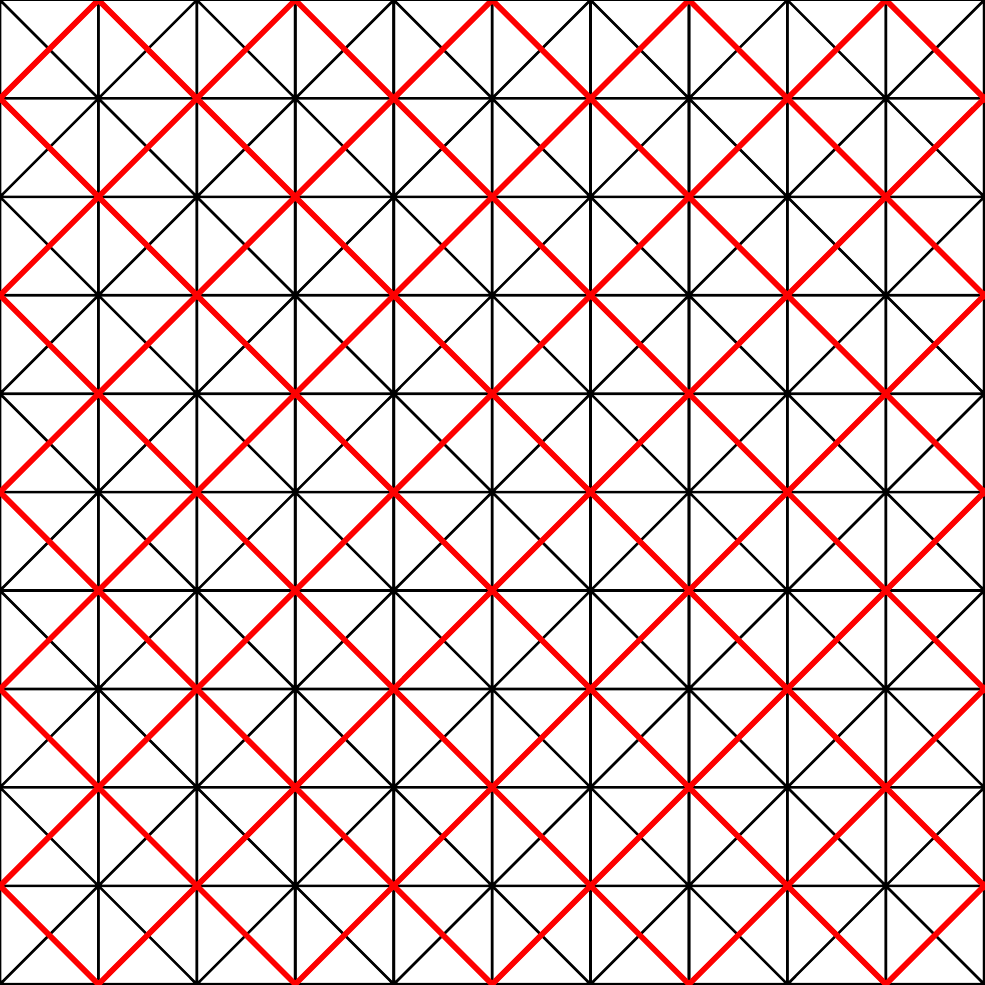}
        \caption{The red edges are an induced subgraph of the Tetrakis grid. They form a square grid.}
        \label{fig:tetrakis_square}
    \end{minipage}
    \end{figure}

    The dual of 4.8.8 is known as the Tetrakis tiling, and contains the square grid as an induced subgraph. In Figure \ref{fig:tetrakis}, the graph induced by the red vertices is a square grid graph. Hamiltonicity in square grid graphs is NP-hard, so Hamiltonicity is also NP-hard in the Tetrakis tiling.
    
    \subsubsection*{Dual of 3.12.12}
\begin{figure}[h]
    \centering
    \begin{minipage}[b]{0.45\linewidth}
        \centering
        \includegraphics[width=5cm]{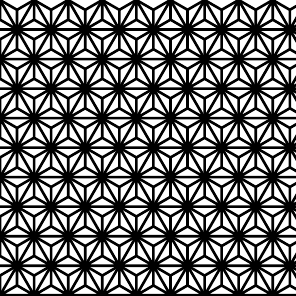}
        \caption{Structure of the dual of 3.12.12.}
        \label{fig:triakis}
        \end{minipage}
    \hspace{0.5cm}
    \begin{minipage}[b]{0.45\linewidth}
        \centering
        \includegraphics[width=5cm]{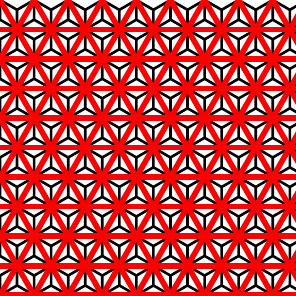}
        \caption{The red edges are an induced subgraph of the dual of 3.12.12, and form a triangular grid.}
        \label{fig:triakis_tri}
    \end{minipage}
    \end{figure}

    The dual of 3.12.12 is known as the Triakis tiling, and contains the triangular grid as an induced subgraph. In Figure \ref{fig:triakis}, the graph induced by the red vertices is a triangular grid graph. Hamiltonicity in triangular grid graphs is NP-hard, so Hamiltonicity is also NP-hard in the Triakis tiling.

    \subsection{Reductions from Hamiltonicity in Hexagonal Grid Graphs}
    \label{sec:Hex}
    In this section, we emulate hexagonal grid graphs via gadgets. For each dual, we construct two kinds of vertex gadgets, which we will call \emph{even} and \emph{odd} vertex gadgets respectively. We will also construct edge gadgets that link even vertex gadgets to odd vertex gadgets. We note that if we can construct gadgets such that every odd vertex gadget can be connected to three even vertex gadgets and vice versa, we will have a three-regular bipartite graph --- this is identical to a hexagonal grid graph, in which Hamiltonicity is known to the NP-Complete. Furthermore, by construction Hamiltonian path through our constructed graph will be possible if and only if a Hamiltonian path exists in the corresponding hexagonal grid. \\ \\
    In order to create gadgets that allow for the creation of a hexagonal grid in which Hamiltonicity is preserved, we impose the following constraints. First, edge gadgets will have the following properties:
    
    \begin{enumerate}
        \item Edge gadgets contain at least one vertex such that the edge will become disconnected if the vertex is removed. This prevents the edge from being used more than once, since a Hamiltonian Cycle using the edge gadget will necessarily pass through this vertex.
        \item Edge gadgets connect to odd vertex gadgets at one vertex and even vertex gadgets at two adjacent vertices.
        \item Edge gadgets have at least two Hamiltonian paths:
        \begin{enumerate}
            \item The ``traverse" path goes from the even side to the odd side.
            \item The ``return" path goes from one vertex of the even side to the other vertex on the even side.
        \end{enumerate}
    \end{enumerate}
    
    \noindent Vertex gadgets will have the following properties:
    
    \begin{enumerate}
        \item As we are emulating hexagonal grids, all vertex gadgets will have 3 entrances/exits.
        \item There is a Hamiltonian path through the vertex gadget starting at any entrance and leaving at any other entrance.
        \item For even vertex gadgets, the Hamiltonian path must also pass through each of the two adjacent vertices connecting it to an edge gadget in succession. This ensures that edge gadgets that correspond to unused edges in the emulated hexagonal gadgets can get ``picked up" by even vertex gadgets.
    \end{enumerate}
    
    The gadgets we present in the following section will obey these properties.
    
    \subsubsection*{Dual of 4.6.12}
\begin{figure}[h]
     \begin{minipage}[b]{0.45\linewidth}
        \centering
        \includegraphics[width=2.5cm]{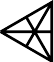}
        \hspace{.5cm}
        \includegraphics[width=2.5cm]{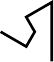}
        \caption{An odd vertex and a path through it, induced by the dual of 4.6.12.}
        \label{fig:4_6_12_odd}
    
    \end{minipage}
    \hspace{0.5cm}
    \begin{minipage}[b]{0.45\linewidth}   
    
        \centering
        \includegraphics[width=3cm]{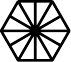}
        \hspace{0.5cm}
        \includegraphics[width=3cm]{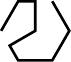}
        \caption{An even vertex and a path through it, induced by the dual of 4.6.12.}
        \label{fig:4_6_12_even}
        \end{minipage}
     \end{figure}

\begin{figure}[h]
        \centering
        \includegraphics[width=4cm]{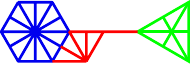}
        \hspace{1cm}
        \includegraphics[width=4cm]{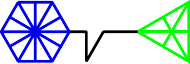}
        \hspace{1cm}
        \includegraphics[width=4cm]{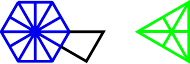}
        \caption{Left: an edge gadget for the dual of 4.6.12, connected to an even vertex on the left and an odd vertex on the right. Middle: A traverse path through the edge gadget. Right: A return path through the edge gadget.}
        \label{fig:4_6_12_edge}
     \end{figure}

\begin{figure}[h]
        \centering
        \includegraphics[width=6cm]{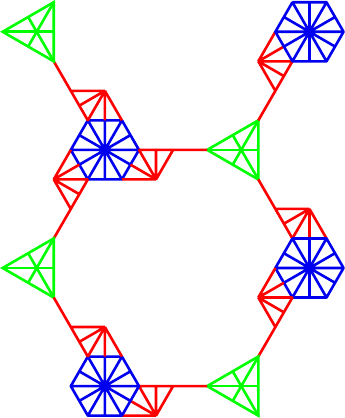}
        \hspace{1cm}
        \includegraphics[width=6cm]{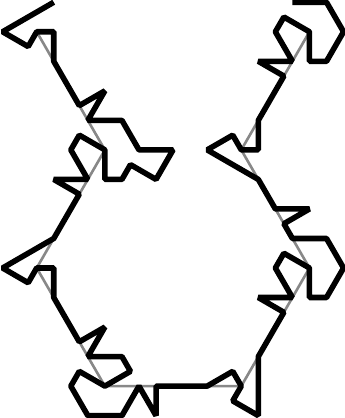}
        \caption{Left: an example of an induced subgraph of the dual of 4.6.12 that emulates a hexagonal grid graph. Right: A Hamiltonian path through said subgraph. The grey lines represent the path through the corresponding hexagonal grid.}
        \label{fig:4_6_12}
    \end{figure}

    In Figure \ref{fig:4_6_12}, green edges are odd vertex gadgets, blue edges are even vertex gadgets, and the red edges are edge gadgets. Figures \ref{fig:4_6_12_odd} and \ref{fig:4_6_12_even} demonstrates paths through even and odd vertices, and \ref{fig:4_6_12_edge} shows traverse and return paths through the edge gadgets. We show that our gadgets satisfy all necessary properties. \\ \\
    First, consider the edge gadget. As shown in Figure \ref{fig:4_6_12_edge} there are both transverse and return paths for this gadget. Furthermore, there is a single vertex that can be removed that will disconnect the odd (green) and even (blue) vertex gadgets that are connected by the edge gadget. Finally, the edge gadget joins to the odd (green) vertex gadget at a single vertex, and to the even (blue) vertex gadget at two vertices. \\ \\
    Next, consider the vertex gadgets. As in Figures \ref{fig:4_6_12_odd} and \ref{fig:4_6_12_even}, it is clear that for both vertex gadgets there are three ways in which an edge gadget can be attached. Furthermore, as depicted for a single possible path, the vertex gadget can only be traversed once by a Hamiltonian path. Lastly, for the even (blue) vertex gadget, as exemplified in Figure \ref{fig:4_6_12_even}, any Hamiltonian path must visit both of the two adjacent vertices connecting it to an edge gadget in succession.
     
    \subsubsection*{Dual of 3.3.3.3.6}
    
\begin{figure}
        \centering
        \includegraphics[width=3.5cm]{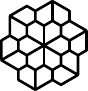}
        \hspace{.5cm}
        \includegraphics[width=3.5cm]{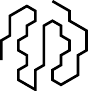}
        \hspace{.5cm}
        \includegraphics[width=3.5cm]{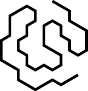}
        \caption{The odd vertex gadget of the dual of 3.3.3.3.6, and two example paths through it.}
        \label{fig:33336_odd}
    \end{figure}

\begin{figure}
        \centering
        \includegraphics[width=3.5cm]{pdf_imgs/33336_dual_odd_vertex_gadget.pdf}
        \hspace{.5cm}
        \includegraphics[width=3.5cm]{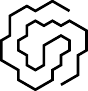}
        \hspace{.5cm}
        \includegraphics[width=3.5cm]{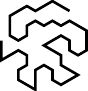}
        \caption{The even vertex gadget of the dual of 3.3.3.3.6, and two example paths through it. Notice that while the shape of the gadget is identical to the odd vertex, the starting and ending points of the paths are different.}
        \label{fig:33336_even}
    \end{figure}

\begin{figure}
        \centering
        \includegraphics[width=5cm]{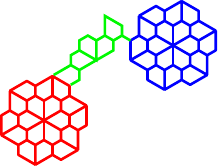}
        \includegraphics[width=5cm]{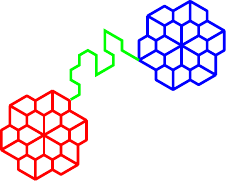}
        \includegraphics[width=5cm]{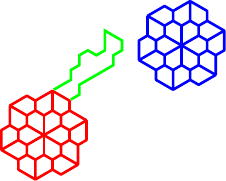}
        \caption{Traverse and return paths through the edge gadgets of the dual of 3.3.3.3.6 are shown in green. We also demonstrate how the edge gadget is attached to even and odd vertices; red edges represent an even vertex and blue edges represent an odd vertex.}
        \label{fig:33336_traverse_return}
    \end{figure}

    Here we present the edge and vertex gadgets for the dual of the 3.3.3.3.6 gadget. Figures \ref{fig:33336_odd} and \ref{fig:33336_even} shows the even and odd vertex gadgets, and example paths through them, while Figure \ref{fig:33336_traverse_return} shows traverse and return paths through the edge gadgets. These gadgets also hold the desired properties to show that Hamiltonicity is NP-Complete.
    
    \subsection{Reductions from Tree-Residue Vertex Breaking}
    \label{sec:TRVB}
    The tree-residue vertex breaking problem asks, given a graph with some ``breakable" vertices, is it possible to break some of them such that the resulting graph is a tree. Demaine and Rudoy \cite{demaine2017tree} characterize the complexity of this problem in planar and bounded degree settings. In particular, they show the problem is still hard in the case of a planar graph with breakable vertices of maximum degree $4$. \\ \\ 
    Our proofs will use a different type of vertex and edge gadget. Vertex gadgets must be able to satisfy ``broken" and ``unbroken" states; furthermore, edge gadgets are simply pairs of paths that have no (or at the very least, no potentially usable) shared edges between them. In each of our proofs below we will focus on the vertex gadgets in broken and unbroken states, as edge gadgets can be any reasonable pairs of paths. We only need to check that the edge gadgets near the vertex do not interfere with each other.
    
    \subsubsection*{Dual of 3.4.6.4}
    
\begin{figure}
    \begin{minipage}[b]{0.52\linewidth}   
        \centering
        \includegraphics[width=8cm]{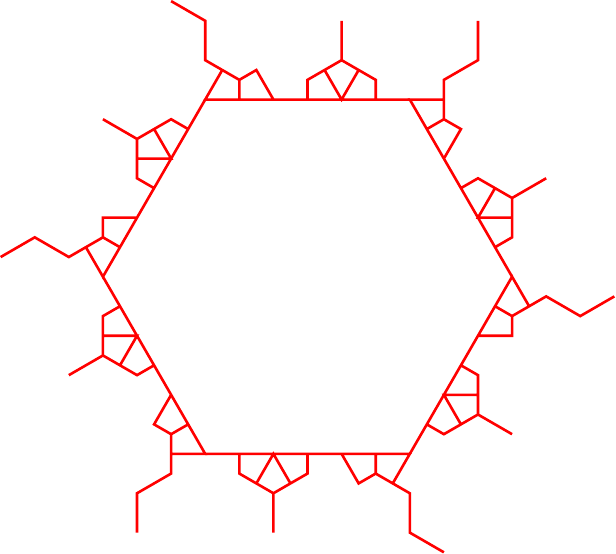}
        \caption{Vertex gadget for the dual of 3.4.6.4.}
        \label{fig:3464_vertex}
    \end{minipage}
    \hspace{.5cm}
    \begin{minipage}[b]{0.38\linewidth}   
        \centering
        \includegraphics[width=6cm]{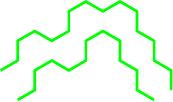}
        \vspace{1.5cm}
        \caption{Example edge gadget for the dual of 3.4.6.4.}
        \label{fig:3464_edge}
    \end{minipage}
    \end{figure}

\begin{figure}
        \centering
        \includegraphics[width=6cm]{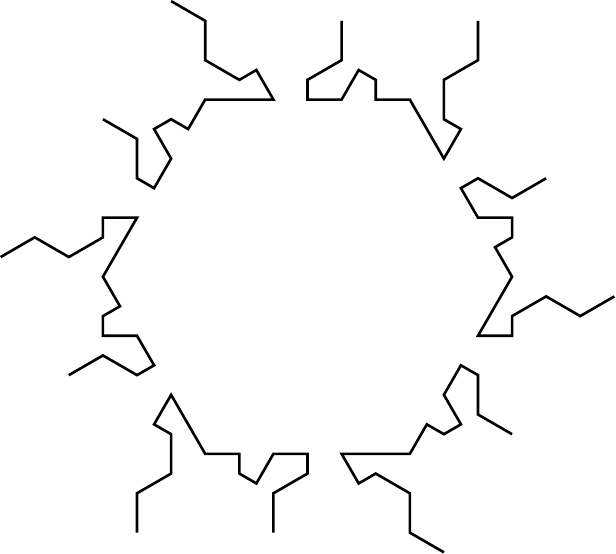}
        \hspace{1cm}
        \includegraphics[width=6cm]{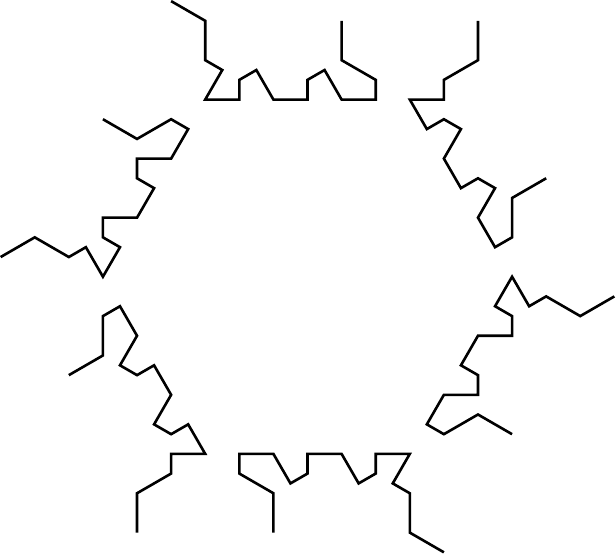}
        \caption{Broken and unbroken states for a vertex gadget for the dual of 3.4.6.4}
        \label{fig:3464_broken_unbroken}
    \end{figure}

    For the dual of 3.4.6.4, we use the 4-regular breakable vertex gadget shown in Figure\ref{fig:3464_vertex}. The broken and unbroken versions are shown in Figure\ref{fig:3464_broken_unbroken}. Edge gadgets can be, for example, pairs of parallel paths extending in the four cardinal directions. The existence of these gadgets proves that Hamiltonicity in the dual of 3.3.4.3.4 is NP-hard.
    
\begin{figure}
        \centering
        \includegraphics[width=6cm]{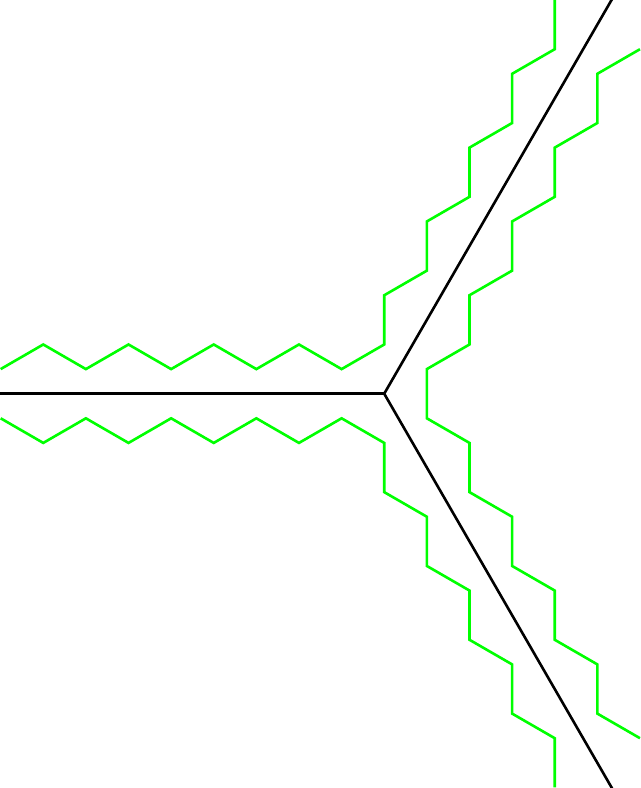}
        \caption{In the dual of 3.4.6.4, three edge gadgets can be joined to emulate an unbreakable degree 3 vertex. The black lines show edges in the graph being emulated.}
        \label{fig:3464_emulate}
    \end{figure}

    A note: it was previously believed that a single degree $6$ breakable vertex gadget was not sufficient to prove NP-hardness, because there is no planar graph with only degree $6$ vertices. However, edge gadgets can modified to create arbitrary degree unbreakable vertex gadgets, so our two gadgets are sufficient. As an example, in figure \ref{fig:3464_emulate}, three edge gadgets in the dual of 3.4.6.4 combine for emulate an unbreakable degree 3 vertex. The existence of this gadget shows that the Hamiltonian path problem in this grid is NP-hard.
    
    \subsubsection*{Dual of 3.3.4.3.4}
    
\begin{figure}
        \centering
        \includegraphics[width=6cm]{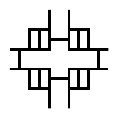}
        \caption{Vertex gadget for the dual of 3.3.4.3.4}
        \label{fig:33434_vertex}
    \end{figure}

\begin{figure}
        \centering
        \includegraphics[width=5cm]{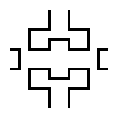}
        \includegraphics[width=5cm]{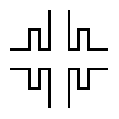}
        \caption{Broken and unbroken states for a 3.3.4.3.4 dual vertex gadget}
        \label{fig:33434_broken_unbroken}
    \end{figure}

    For the dual of 3.3.4.3.4, we use the 4-regular breakable vertex gadget shown in Figure~\ref{fig:33434_vertex}. The breakable and unbreakable versions are shown in Figure~\ref{fig:33434_broken_unbroken}. Edge gadgets can be, for example, pairs of parallel paths extending in the four cardinal directions. The existence of these gadgets proves that Hamiltonicity in the dual of 3.3.4.3.4 is NP-hard.

    \subsubsection*{Dual of 3.3.3.4.4}
    
\begin{figure}
        \centering
        \includegraphics[width=4cm]{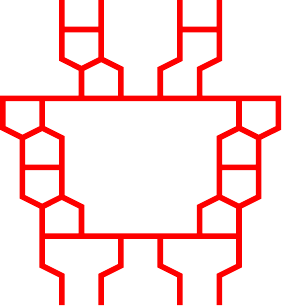}
        \hspace{.5cm}
        \includegraphics[width=4cm]{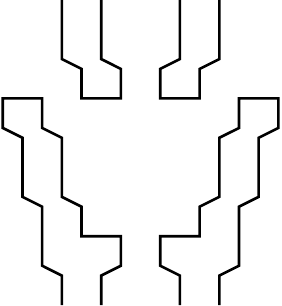}
        \hspace{.5cm}
        \includegraphics[width=4cm]{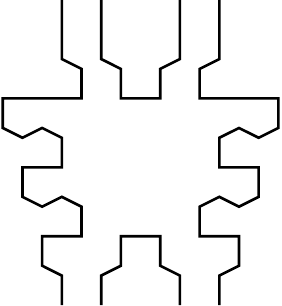}
        \caption{Left: vertex gadget for the dual of 3.3.3.4.4. Middle and right: broken and unbroken states for said gadget.}
        \label{fig:33344_broken_unbroken}
    \end{figure}

    For the dual of 3.3.3.4.4, we use the 4-regular breakable vertex gadget shown in Figure \ref{fig:33344_broken_unbroken}, with the corresponding broken and unbroken versions shown in the red edges. The full gadget is simply the union of the red and blue edges. The existence of this vertex gadget proves that Hamiltonicity in the dual of 3.3.3.4.4 is NP-hard. With this, we have that the Hamiltonian path problem is NP-hard in the duals of all eight semi-regular tessellations.

    \subsubsection{Augmented square grids}
    In this section we show that Hamiltonicity in augmented square grids is NP-complete. Further, our reduction works for thin graphs, giving alternate proofs and sometimes stronger results for a number of well known grid graphs.
    
 A \emph{full augmented square grid graph} is an infinite graph formed by taking the infinite square grid graph and adding either or both diagonals of each square pixel may exist as edges in the graph. An \emph{augmented square grid graph} is an induced subgraph of a full augmented square grid graph. We will show that the Hamiltonian Cycle Problem for any augmented square grid is NP-hard. As before, we will use regular degree-4 tree-residue vertex breaking.\\ \\ 
 
\begin{figure}
        \centering
        \begin{minipage}[b]{.45\linewidth}
        \centering
        \includegraphics[width=6cm]{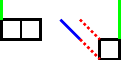}
        \caption{Edges to consider for gadget creation in augmented grid graphs. The green edges denote a possible edge gadget.}
        \label{fig:aug_start}
        \end{minipage}
        \hspace{.5cm}
        \begin{minipage}[b]{.45\linewidth}
        \centering
        \includegraphics[width=6cm]{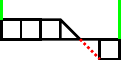}
        \caption{Case 1: connection used when the blue edge exists. The red dashed line may or may not be induced; the gadget works either way.}
        \label{fig:aug_1}
        \end{minipage}
        
        \vspace{.3cm}
        \begin{minipage}[b]{.45\linewidth}
        \centering
        \includegraphics[width=6cm]{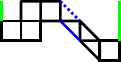}
        \caption{Case 2: connection used when the solid blue edge doesn't exist but both red edges do. The dotted blue edge may or may not exist, the gadget works either way.}
        \label{fig:aug_2}
        \end{minipage}
        \hspace{.5cm}
        \begin{minipage}[b]{.45\linewidth}
        \centering
        \includegraphics[width=6cm]{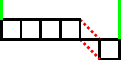}
        \vspace{.25cm}
        \caption{Case 3: When at least one red edge does not exist. The gadget still works if only one of them exists.}
        \label{fig:aug_3}
        \end{minipage}
    \end{figure}

    Many interesting families of graphs are augmented square grids. These include: the King's graph, the box-pleat grid (important in origami design), the triangular grid graph, the 3.3.4.3.4 and 3.3.3.4.4 semi-regular tessellations, and other combinations of unit right triangles and squares. It is also interesting to note that this includes aperiodic tilings of the plane, for example by including the diagonals in every pixel whose bottom-leftmost point has coordinates that sum to powers of 10. \\ \\    Edge gadgets are simple, we can just use arbitrary parallel paths that are distance $7$ apart. All vertex gadgets will be contained in a $16$ by $16$ square, and will contain 4 ``side" pieces and 4 ``corner" pieces. Each side piece will connected to exactly 2 corner pieces and each corner piece will be connected to exactly 2 side pieces; the final construction will look somewhat like a square. We will focus on a $3 \times 6$ region that amounts to one eighth of the final vertex gadget, which is focused on the connection between a side and corner piece; we will demonstrate how to combine the regions into a full vertex gadget after all of our casework. Our diagrams will concentrate on the right half of the top side of the vertex gadget. \\ \\ 
    Consider Figure \ref{fig:aug_start}. All the black edges will be guaranteed to be part of our gadget. We now have three cases. If the blue edge exists, we use the connection in Figure \ref{fig:aug_1}. If the blue edge does not exist but \emph{both} red ones do, we use the connection in Figure \ref{fig:aug_2}. Finally, if either red edge does not exist, we use the connection in Figure \ref{fig:aug_3}.
    
    In each of the three cases we described, we have only drawn the minimum number of edges in the induced subgraph (there may exist other induced diagonal edges in the square pixels. However, notice that in each connection there exists a vertex such that, if that vertex is removed, the gadget becomes separated. This is essential, because it means the connection can be traversed at most one time. \\ \\
    
\begin{figure}
    \centering
        \begin{minipage}[b]{0.55\linewidth}
        \centering
        \includegraphics[width=7.5cm]{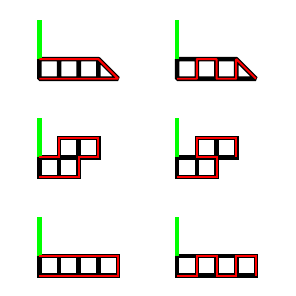}
        \caption{Demonstration of a Hamiltonian path through a side piece. Combine one half side piece on the left with one reflected half side piece on the right for a path through a full side gadget. See Figure \ref{fig:aug_combo} for a few examples.}
        \label{fig:aug_path_side}
        \end{minipage}
        \hspace{.4cm}
        \begin{minipage}[b]{0.4\linewidth}
        \centering
        \includegraphics[width=5cm]{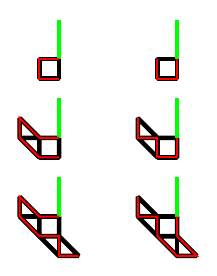}
        \vspace{.5cm}
        \caption{Demonstration of a Hamiltonian path through a corner piece; all corner pieces are one of these three cases.}
        \label{fig:aug_path_corner}
        \end{minipage}
    \end{figure}

    Furthermore, we need to verify that for each side or corner piece, it is possible to create a Hamiltonian path through each edge/corner piece, with all combinations of ``half pieces," that starts at the edge gadget and ends at the connection point. The case for side and corner pieces is shown in figures \ref{fig:aug_path_side} and \ref{fig:aug_path_corner}.
    
\begin{figure}
        \centering  
        \includegraphics[width=8cm]{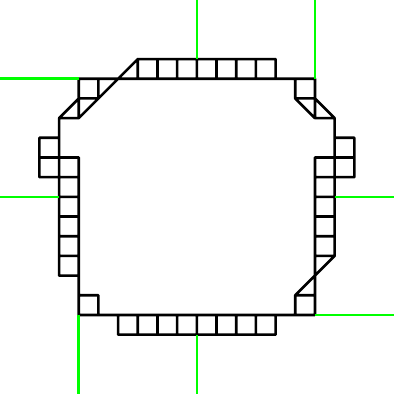} \\
        \begin{minipage}[b]{0.45\linewidth}
        \includegraphics[width=6cm]{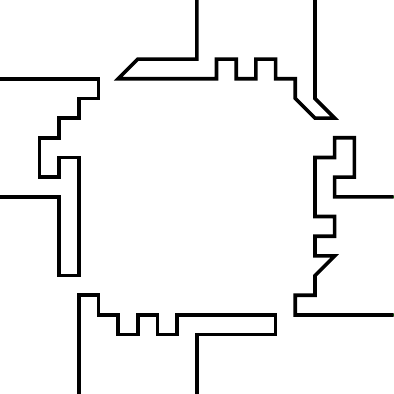}
        \end{minipage}
        \hspace{.5cm}
        \begin{minipage}[b]{0.45\linewidth}
        \includegraphics[width=6cm]{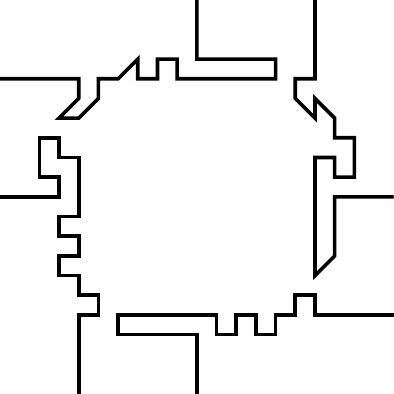}
        \end{minipage}
        \caption{Top: A full vertex gadget combining multiple types of connections. Bottom: Broken and unbroken states for this particular gadget.}
        \label{fig:aug_combo}
    \end{figure}

    As an example of what a complete gadget, see Figure \ref{fig:aug_combo}. This demonstrates combining eight different gadgets to create a breakable degree-4 vertex gadget. Green lines represent edge gadgets, which are relatively unconstrained. It is straightforward to verify that any Hamiltonian path or cycle involving this vertex gadget forces it into either the broken or unbroken states. The existence of this gadget proves that the Hamiltonian path problem is NP-hard in all augmented square grid graphs.

    \section{Conclusion and Further Work}
    In this paper, we have shown that the Hamiltonian path problem in many tessellations is NP-Hard. Using reductions to Hamiltonicity in other grid graphs, we have shown that Hamiltonicity in the duals of the 3.6.3.6, 4.8.8, 3.12.12, 4.6.12, and 3.3.3.3.6 tessellations is NP-hard. Furthermore, using reductions to tree-residue vertex breaking, we have shown Hamiltonicity in the duals of the 3.4.6.4, 3.3.4.3.4, and 3.3.3.4.4 tessellations, as well as augmented square grid graphs, is NP-hard. \\ \\

    A wide variety of tessellations can be categorized as certain augmented square grid graph, so those tessellations must also be NP-hard. This leads to alternative proofs for Hamiltonicity in e.g. the triangular, hexagonal, 3.3.3.4.4, and 3.3.4.3.4 grids. Overall, the frameworks we have set up for constructing gadgets to convert Hamiltonicity in various grid graphs to Hamiltonicity in a simple regular tiling or to TRVB will be powerful ways to simplify other open problems in this area.\\ \\
    There are still some questions that remain to be answered. A natural extension to augmented square grids are augmented hexagonal grids. The substantial blow-up in cases for the hexagonal grid makes this more difficult to tackle. Perhaps restricting the augmentation to single edges in each pixel will still provide interesting results but remain more tractable.\\
    
    Although we can show Hamiltonicity in some aperidic tilings is hard, it would be nice to see this shown for a natural one such as the Penrose tiling, Truchet tiling, or Conway's pinwheel tiling. These tessellations may require some new techniques, as it seems somewhat problematic to tackle with techniques we have explored here.\\
    
    Finally, one could examine the thin, solid, and polygonal cases of the semi-regular tessellations and their duals.
    
    \section*{Acknowledgements}
    This paper came out of the MIT course 6.892 Algorithmic Lower Bounds: Fun with Hardness Proofs taught by Professor Erik Demaine. In addition to Professor Demaine, we would like to thank the students and teaching staff for their support and ideas. We would also like to thank Kaiying Hou for some discussion and ideas related to this paper.
    
\bibliography{ref}{}
\bibliographystyle{alpha}

\end{document}